\documentclass{amsart}
\usepackage{amssymb,amsmath}
\usepackage{amsopn}
\usepackage{mathrsfs}
\usepackage{color}
\usepackage{wasysym}
\usepackage{vmargin}
\usepackage{amscd}
\usepackage{verbatim}

\newcommand\RR{{{\mathbb R}}}

\newcommand\SSS{{\mathbb S}}
\newcommand{\rr}{\mathbb{R}}
\newcommand{\eps}{\varepsilon}
\newcommand{\nn}{\mathbb{N}}

\def\un{{\mathrm{1~\hspace{-1.4ex}l}}}
\def\triple{|\hspace{-0.4mm}|\hspace{-0.4mm}|}

\def\Id{\operatorname{Id}}

\newtheorem{theorem}{Theorem}[section]
\newtheorem{lemma}[theorem]{Lemma}
\newtheorem{proposition}[theorem]{Proposition}

\numberwithin{equation}{section}
\def\p{\partial}

\def\io{{\infty}}

\def\Id{\operatorname{Id}}

\def\R{\mathbb R}
\def\C{\mathbb C}
\def\poscal#1#2{\langle#1,#2\rangle}

\def\norm#1{\Vert#1\Vert}
\def\val#1{\vert#1\vert}

\def\valjp#1{\langle#1\rangle}

\def\l2{L^2(\R^{n})}
\def\L2{L^2(\R^{2n})}

\def\vs{\vskip.3cm}

\def\mat22#1#2#3#4{\begin{pmatrix}#1&#2\\ #3&#4\end{pmatrix}}

\def\finp{\operatorname{fp}}

\begin{document}

\title[On the linearized Landau and Boltzmann operators]
{Phase space analysis and functional calculus for the linearized Landau and Boltzmann operators}

\author{N. Lerner, Y. Morimoto, K. Pravda-Starov  \& C.-J. Xu\\{}}
\date{\today}
\address{\noindent \textsc{N. Lerner, Institut de Math\'ematiques de Jussieu,
Universit\'e Pierre et Marie Curie (Paris VI),
4 Place Jussieu,
75252 Paris cedex 05,
France}}
\email{lerner@math.jussieu.fr}
\address{\noindent \textsc{Y. Morimoto, Graduate School of Human and Environmental Studies,
Kyoto University, Kyoto 606-8501, Japan}}
\email{morimoto@math.h.kyoto-u.ac.jp }
\address{\noindent \textsc{K. Pravda-Starov,
Universit\'e de Cergy-Pontoise,
CNRS UMR 8088,
D\'epartement de Math\'ematiques,
95000 Cergy-Pontoise, France}}
\email{karel.pravda-starov@u-cergy.fr}
\address{\noindent \textsc{C.-J. Xu, Universit\'e de Rouen, CNRS UMR 6085, Laboratoire de Math\'ematiques, 76801 Saint-Etienne du Rouvray, France\\ and \\
School of Mathematics, Wuhan university 430072, Wuhan, P.R. China}}
\email{Chao-Jiang.Xu@univ-rouen.fr}

\begin{abstract}
In many works, the linearized non-cutoff Boltzmann operator is considered to behave essentially as a fractional Laplacian. 
In the present work, we prove that the linearized non-cutoff Boltzmann operator with Maxwellian molecules is exactly equal to a fractional power of the linearized Landau operator which is the sum of the harmonic oscillator and the spherical Laplacian.  This result allows to display explicit sharp coercive estimates satisfied by the linearized non-cutoff Boltzmann operator for both Maxwellian and non-Maxwellian molecules.
\end{abstract}

\maketitle

\section{Introduction}
The Boltzmann equation describes the behaviour of a dilute gas when the only interactions taken into account are binary collisions \cite{17}. It reads as the equation
\begin{equation}\label{e1}
\begin{cases}
\partial_tf+v\cdot\nabla_{x}f=Q_B(f,f),\\
f|_{t=0}=f_0,
\end{cases}
\end{equation}
for the density distribution of the particles in the gas $f=f(t,x,v) \geq 0$  at time $t$, having position $x \in \rr^d$ and velocity $v \in \rr^d$. The Boltzmann equation derived in 1872 is one of the fundamental equations of mathematical physics and, in particular, a cornerstone of statistical physics.

The term appearing in the right-hand-side of this equation $Q_B(f,f)$ is the so-called Boltzmann collision operator associated to the Boltzmann bilinear operator
$$
Q_B(g, f)=\int_{\rr^d}\int_{\SSS^{d-1}}B(v-v_{*},\sigma) \big(g'_* f'-g_*f\big)d\sigma dv_*,
$$
with $d \geq 2$, where we are using the standard shorthand $f'_*=f(t,x,v'_*)$, $f'=f(t,x,v')$, $f_*=f(t,x,v_*)$, $f=f(t,x,v)$. In this expression, $v$, $v_*$ and $v'$, $v_*'$ are the velocities in $\rr^d$ of a pair of particles before and after the collision. They are connected through the formulas
$$
v'=\frac{v+v_*}{2}+\frac{|v-v_*|}{2}\sigma,\quad   v_*'=\frac{v+v_*}{2}-\frac{|v-v_*|}{2}\sigma,
$$
where $\sigma\in\SSS^{d-1}$. Those relations correspond physically to elastic collisions with the conservations of momentum and kinetic energy in the binary collisions
$$\quad v+v_{\ast}=v'+v_{\ast}', \quad |v|^2+|v_{\ast}|^2=|v'|^2+|v_{\ast}'|^2,$$
where $|\cdot|$ is the Euclidean norm on $\rr^d$. In the present work, our main focus is to study the sharp anisotropic diffusive effects induced by this operator under general physical assumptions on the collision kernel.

For monatomic gas, the cross section $B(v-v_{*},\sigma)$ is a non-negative function which only depends on the relative velocity $|v-v_*|$ and on the deviation angle $\theta$ defined through the scalar product in~$\rr^d$,
$$\cos \theta= k \cdot \sigma, \qquad  k=\frac{v-v_*}{|v-v_*|}.$$
Without loss of generality, we may assume that $B(v-v_{*},\sigma)$ is supported on the set where
$$k \cdot \sigma \geq 0,$$
i.e. where $0 \leq \theta \leq \frac{\pi}{2}$. Otherwise, we can reduce to this situation with the customary symmetrization
$$\tilde{B}(v-v_{*},\sigma)=\big[B(v-v_{*},\sigma)+B(v-v_{*},-\sigma)\big] \un_{\{\sigma \cdot k \geq 0\}},$$
with $\un_A$ being the characteristic function of the set $A$, since the term $f'f_*'$ appearing in the Boltzmann operator $Q_B(f,f)$ is invariant under the mapping $\sigma \rightarrow -\sigma$.
More specifically, we consider cross sections of  the type
$$
B(v-v_*,\sigma)=\Phi(|v-v_*|)b\Big(\frac{v-v_*}{|v-v_*|} \cdot \sigma\Big),
$$
with a kinetic factor
\begin{equation}\label{sa0}
\Phi(|v-v_*|)=|v-v_*|^{\gamma}, \quad  \gamma \in ]-d,+\infty[,
\end{equation}
and a factor related to the deviation angle with a singularity
\begin{equation}\label{sa1b}
(\sin \theta)^{d-2}b(\cos \theta)  \substack{\\ \\ \approx \\ \theta \to 0_{+} }  \theta^{-1-2s},
\end{equation}
for\footnote{The notation $a\approx b$ means $a/b$
is bounded from above and below by fixed positive constants.} some  $0 < s <1$. Notice that this singularity is not integrable
$$\int_0^{\frac{\pi}{2}}(\sin \theta)^{d-2}b(\cos \theta)d\theta=+\infty.$$
This non-integrability property plays a major r\^ole regarding the qualitative behaviour of the solutions of the Boltzmann equation and this non-integrability feature is essential for the smoothing effect to be present. Indeed, as first observed by Desvillettes for the Kac equation in ~\cite{D95}, grazing collisions that account for the non-integrability of the angular factor near $\theta=0$
do induce smoothing effects for the solutions of the non-cutoff Kac equation, or more generally for the solutions of the non-cutoff Boltzmann equation. On the other hand, these solutions are at most as regular as the initial data, see e.g. \cite{36}, when the collision cross section is assumed to be integrable, or after removing the singularity by using a cutoff function (Grad's angular cutoff assumption).

The physical motivation for considering this specific structure of cross sections is derived from particles interacting according to a spherical intermolecular repulsive potential of the form
$$\phi(\rho)=\rho^{-r}, \quad r>1,$$
with $\rho$ being the distance between two interacting particles. In the physical 3-dimensional space~$\rr^3$, the cross section satisfies the above assumptions with $s=\frac{1}{r} \in ]0, 1[$ and $\gamma=1-4s \in ]-3, 1[$. For Coulomb potential $r=1$, i.e. $s=1$, the Boltzmann operator is not well defined \cite{villani3}. In this case, the Landau operator is substituted to the Boltzmann operator \cite{villani2} in the equation (\ref{e1}). The Landau equation was first written by Landau in 1936 \cite{landau}. It is similar to the Boltzmann equation
\begin{equation}\label{e1bis}
\begin{cases}
\partial_tf+v\cdot\nabla_{x}f=Q_L(f,f),\\
f|_{t=0}=f_0,
\end{cases}
\end{equation}
with a different collision operator $Q_L$. Indeed, in the case of long-distance interactions, collisions occur mostly for grazing collisions. When all collisions become concentrated near $\theta=0$, one obtains by the grazing collision limit asymptotic \cite{mou5,mou1,mou2,mou3,villani1} the Landau collision operator
$$
Q_L(g, f)=\nabla_v \cdot \Big(\int_{\RR^d}a(v-v_*)\big(g(t,x,v_*)(\nabla_v f)(t,x,v)-(\nabla_v g)(t,x,v_*)f(t,x,v)\big)d v_*\Big),
$$
where $a=(a_{i,j})_{1 \leq i,j \leq d}$ stands for the non-negative symmetric matrix
$$
a(v)=(|v|^2\Id -v\otimes v)|v|^{\gamma} \in M_d(\rr), \quad -d<\gamma<+\infty.
$$
The Landau operator is understood as the limiting Boltzmann operator in the case when $s=1$ in the singularity assumption (\ref{sa1b}). We shall confirm this feature and prove that for Maxwellian molecules, the linearized non-cutoff Boltzmann operator is truly equal to a fractional linearized Landau operator with exponent exactly given by the singularity parameter $0<s<1$.

We shall study the linearizations of the Boltzmann and Landau equations (\ref{e1}), (\ref{e1bis}) by considering the fluctuation
$$f=\mu+\sqrt{\mu}g,$$
around the Maxwellian equilibrium distribution
\begin{equation}\label{maxwe}
\mu(v)=(2\pi)^{-\frac{d}{2}}e^{-\frac{|v|^2}{2}}.
\end{equation}
Since $Q_J(\mu,\mu)=0$, for $J=B$ or $J=L$, by the conservation of the kinetic energy for the Boltzmann operator and a direct computation for the Landau operator, the collision operator $Q_J(f,f)$ can be split into three terms
$$Q_J(\mu+\sqrt{\mu}g,\mu+\sqrt{\mu}g)=Q_J(\mu,\sqrt{\mu}g)+Q_J(\sqrt{\mu}g,\mu)+Q_J(\sqrt{\mu}g,\sqrt{\mu}g),$$
whose linearized part is
$Q_J(\mu,\sqrt{\mu}g)+Q_J(\sqrt{\mu}g,\mu).$
Setting
$$\mathscr{L}_Jg=\mathscr{L}_{1,J}g+\mathscr{L}_{2,J}g,$$
with
$$\mathscr{L}_{1,J}g=-\mu^{-1/2}Q_J(\mu,\mu^{1/2}g), \quad \mathscr{L}_{2,J}g=-\mu^{-1/2}Q_J(\mu^{1/2}g,\mu),$$
the original Boltzmann and Landau equations (\ref{e1}), (\ref{e1bis}) are reduced to the Cauchy problem for the fluctuation
$$
\begin{cases}
\partial_tg+v\cdot\nabla_{x}g+\mathscr{L}_Jg=\mu^{-1/2}Q_J(\sqrt{\mu}g,\sqrt{\mu}g),\\
g|_{t=0}=g_0.
\end{cases}
$$
These collision operators are local in the time and position variables and from now on, we consider them as acting
only in the velocity variable.
These linearized operators $\mathscr{L}_B$, $\mathscr{L}_L$ are known \cite{17,mou7,Guo1,mou6} to be unbounded symmetric operators on $L^2(\rr^d_{v})$ (acting in the velocity variable) such that their Dirichlet form satisfy
$$(\mathscr{L}_Bg,g)_{L^2(\rr^d_{v})} \geq 0, \qquad (\mathscr{L}_Lg,g)_{L^2(\rr^d_{v})} \geq 0.$$
Setting
$$\mathbf{P}g=(a+b \cdot v+c|v|^2)\mu^{1/2},$$
with $a,c \in \rr$, $b \in \rr^d$, the $L^2$-orthogonal projection onto the space of collisional invariants
\begin{equation}\label{coli}
\mathcal{N}=\textrm{Span}\big\{\mu^{1/2},v_1 \mu^{1/2},...,v_d\mu^{1/2},|v|^2\mu^{1/2}\big\},
\end{equation}
we have
\begin{equation}\label{ker}
(\mathscr{L}_Bg,g)_{L^2(\rr^d)}=0 \Leftrightarrow g=\mathbf{P}g, \qquad (\mathscr{L}_Lg,g)_{L^2(\rr^d)}=0 \Leftrightarrow g=\mathbf{P}g.
\end{equation}
It was noticed forty years ago by Cercignani \cite{cercignani}  that the linearized Boltzmann operator $\mathscr{L}_B$ with Maxwellian molecules, i.e. when the parameter $\gamma=0$ in (\ref{sa0}), behaves like a fractional diffusive operator. Over the time, this point of view transformed into the following widespread heuristic conjecture on the diffusive behavior of the Boltzmann collision operator as a flat fractional Laplacian \cite{al2009,al-1,amuxy-2,pao1,pao2,villani2}:
$$f \mapsto Q_B(\mu,f) \sim -(-\Delta_v)^sf+ \textrm{ lower order terms},$$
with $0<s<1$ being the parameter appearing in the singularity assumption (\ref{sa1b}). See \cite{lmp,MoXu,M-X2} for works related to this simplified model of the non-cutoff Boltzmann equation. Regarding the general non-cutoff linearized Boltzmann operator, sharp coercive estimates in the weighted isotropic Sobolev spaces $H^k_l(\rr^d)$ were proven in \cite{amuxy3,amuxy-4-1,gr-st,44,strain}:
\begin{equation}\label{sa2b}
 \left\|
(1 - {\bf P} ) g\right\|^2_{H^s_{\frac{\gamma}{2}}}+\left\|
(1 - {\bf P} ) g\right\|^2_{L^2_{s+\frac{\gamma}{2}}}
\lesssim (\mathscr L_B g,\, g)_{L^2(\rr^d)} \lesssim \left\|
(1 - {\bf P} )g\right\|^2_{H^s_{s+\frac{\gamma}{2}}},
\end{equation}
where
$$H^k_l(\rr^d) = \big\{ f\in \mathscr S ' (\RR^d ) : \ (1+|v|^2)^{\frac{l}{2}} f \in H^k (\RR^d ) \big\}, \quad k, l \in \RR.$$
In the recent work \cite{LPMX1}, we investigate the exact phase space structure of the linearized non-cutoff Boltzmann operator with Maxwellian molecules acting on radially symmetric functions with respect to the velocity variable. This linearized non-cutoff radially symmetric Boltzmann operator was shown to be exactly an explicit function of the harmonic oscillator
$$
\mathcal{H}=-\Delta_v+\frac{|v|^2}{4}.
$$
It is diagonal in the Hermite basis and behaves essentially as the fractional harmonic oscillator
$$\Big(1-\Delta_v+\frac{|v|^2}{4}\Big)^s,$$
where $0<s<1$ is the parameter appearing in the singularity assumption \eqref{sa1b}. This linearized operator was also studied from a microlocal view point and shown to be a pseudodifferential operator
$$\mathscr{L}_Bf=l^w(v,D_v)f,$$
when acting on radially symmetric Schwartz functions $f  \in \mathscr{S}_r(\rr_v^d)$, whose symbol belongs to a standard symbol class and admit a complete asymptotic expansion
$$
l(v,\xi) \sim
c_0\Big(1+\vert\xi|^2+\frac{|v|^2}{4}\Big)^{s}-d_{0}
+\sum_{k=1}^{+\infty}c_k\Big(1+\vert\xi|^2+\frac{|v|^2}{4}\Big)^{s-k}, \quad c_0,d_0>0,\  c_k \in \rr,\ k\geq 1.
$$
This asymptotic expansion provides a complete description of the phase space structure of the linearized non-cutoff radially symmetric Boltzmann operator and allows to strengthen in the radially symmetric case with Maxwellian molecules the coercive estimate (\ref{sa2b}) as
$$\|\mathcal{H}^{\frac{s}{2}}(1-{\bf P})f\|_{L^2}^2  \lesssim (\mathscr{L}_Bf,f)_{L^2} \lesssim \|\mathcal{H}^{\frac{s}{2}}(1-{\bf P})f\|_{L^2}^2 , \quad f \in \mathscr{S}_{r}(\rr^d),$$
where $\mathcal H$ is the harmonic oscillator. However, the general (non radially symmetric) Boltzmann operator is a truly anisotropic operator. This accounts in general for the difference between the lower and upper bounds in the sharp estimate (\ref{sa2b}) with usual weighted Sobolev norms. In the recent works \cite{amuxy-4-1,gr-st,gr-st1}, sharp coercive estimates for the general linearized non-cutoff Boltzmann operator were proven. In \cite{amuxy-4-1}, these sharp coercive estimates established in the three-dimensional setting $d=3$ (Theorem~1.1 in \cite{amuxy-4-1}),
\begin{equation}\label{tripleest}
\forall f \in \mathscr{S}(\rr^3), \quad   \triple (1-{\bf P})f\triple_{\gamma}^2  \lesssim (\mathscr{L}_Bf,f)_{L^2} \lesssim  \triple (1-{\bf P})f\triple_{\gamma}^2,
\end{equation}
involve the anisotropic norm
$$
\triple f \triple_{\gamma}^2=\int_{\rr^3_{v} \times \rr^3_{v_*} \times \SSS^2_{\sigma}}|v-v_*|^{\gamma}b(\cos \theta)\big(\mu_*(f'-f)^2+f_*^2(\sqrt{\mu'}-\sqrt{\mu})^2\big)dvdv_*d\sigma,
$$
whereas in \cite{gr-st,gr-st1}, coercive estimates involving the anisotropic norms
$$\|f\|_{N^{s,\gamma}}^2=\|f\|_{L_{\gamma+2s}^2}^2+\int_{\rr^{d}}\int_{\rr^{d}}\langle v \rangle^{\frac{\gamma+2s+1}{2}} \langle v' \rangle^{\frac{\gamma+2s+1}{2}}\frac{|f(v)-f(v')|^2}{d(v,v')^{d+2s}}\un_{d(v,v') \leq 1}dv dv',$$
where
$$d(v,v')=\sqrt{|v-v'|^2+\frac{1}{4}(|v|^2-|v'|^2)^2},$$
were derived and a model of a fractional geometric Laplacian with the geometry of a lifted paraboloid in $\rr^{d+1}$ was suggested for interpreting the anisotropic diffusive properties of the Boltzmann collision operator.

In the present work, we shall prove that in the physical 3-dimensional space the non-cutoff linearized Boltzmann operator with Maxwellian molecules $\mathscr{L}_B$ is actually given by the fractional power of the linearized Landau operator $\mathscr{L}_L^s$. Furthermore, we shall provide more explicit coercive estimates satisfied by the linearized non-cutoff Boltzmann operator for both Maxwellian and non-Maxwellian molecules.

\section{Statements of the main results}

We consider the Landau operator with Maxwellian molecules
$$
Q_L(g, f)=\nabla_v \cdot \Big(\int_{\RR^d}a(v-v_*)\big(g(v_*)(\nabla f)(v)-(\nabla g)(v_*)f(v)\big)d v_*\Big),
$$
where $a=(a_{i,j})_{1 \leq i,j \leq d}$ stands for the non-negative symmetric matrix
$$
a(v)=|v|^2\Id -v\otimes v \in M_d(\rr).
$$
We shall use the following notations. The standard Hermite functions $(\phi_{n})_{n\in \nn}$ are defined on $\rr$ by
 \begin{align*}
 \phi_{n}(x)=& \ (-1)^n(2^n n!)^{-\frac{1}{2}}\pi^{-\frac{1}{4}} e^{\frac{x^2}{2}}\frac{d^n}{dx^n}(e^{-x^2})
\\ = & \
(2^n n!)^{-\frac{1}{2}}\pi^{-\frac{1}{4}} \Big(x-\frac{d}{dx}\Big)^n(e^{-\frac{x^2}{2}})=
(n!)^{-\frac{1}{2}} a_{+}^n \phi_{0},
\end{align*}
where $a_{+}$ is the creation operator $2^{-\frac{1}{2}}(x-\frac{d}{dx})$.
The $(\phi_{n})_{n\in \nn}$ make an orthonormal basis of $L^2(\rr)$.
We denote for $n\in \nn$, $\alpha=(\alpha_{j})_{1\leq j\leq d}\in\nn^d$, $x\in \rr$, $v\in \rr^d,$
\begin{align*}
\psi_n(x)&=2^{-\frac{1}{4}}\phi_n(2^{-\frac{1}{2}}x),\quad \psi_{n}=(n!)^{-\frac{1}{2}}
\Big(\frac{x}2-\frac{d}{dx}\Big)^n\psi_{0},
\\
\Psi_{\alpha}(v)&=\prod_{j=1}^d\psi_{\alpha_j}(v_j),\quad \mathcal E_{k}=\textrm{Span}
\{\Psi_{\alpha}\}_{\alpha\in \nn^d, |\alpha|=k},
\end{align*}
with $|\alpha|=\alpha_{1}+\dots+\alpha_{d}$.
The $(\Psi_{\alpha})_{\alpha \in \nn^d}$
make an orthonormal basis of $L^2(\rr^d)$
composed by the eigenfunctions of the $d$-dimensional harmonic oscillator:
$$
\mathcal{H}=-\Delta_v+\frac{|v|^2}{4}=
\sum_{k\geq 0}\Big(\frac d2+k\Big)\mathbb P_{k},\quad
\text{Id}=\sum_{k \ge 0}\mathbb P_{k},
$$
where
$\mathbb P_{k}$
is the orthogonal projection onto $\mathcal E_{k}$
(whose dimension is $\binom{k+d-1}{d-1}$).
The eigenvalue
$\frac{d}{2}$ is simple in all dimensions and $\mathcal E_{0}$ is generated by
\begin{equation}\label{k2.0}
\Psi_{0}(v)=(2\pi)^{-\frac{d}{4}}e^{- \frac{|v|^2}{4}}=\mu^{1/2}(v),
\end{equation}
with $\mu$ the Maxwellian distribution (\ref{maxwe}).
Notice that for any $1 \leq j,k \leq d$ with $j \neq k$,
\begin{equation}\label{k2.1}
\Psi_{e_k}(v)=v_k \Psi_0(v), \quad \Psi_{2e_k}(v)=\frac{1}{\sqrt{2}}(v_k^2-1) \Psi_0(v), \quad \Psi_{e_j+e_k}(v)=v_j v_k \Psi_0(v),
\end{equation}
if $(e_k)_{1 \leq k \leq d}$ stands for the canonical basis of $\rr^d$. Those formulas show that the space of collisional invariants (\ref{coli}) may be expressed through the Hermite basis as
$$\mathcal{N}=\textrm{Span}\Big\{\Psi_0,\Psi_{e_1},...,\Psi_{e_d},\sum_{j=1}^d\Psi_{2e_j}\Big\}.$$
Our first result which is probably well-known provides an explicit expression for the linearized Landau operator with Maxwellian molecules:

\bigskip

\begin{proposition}\label{th1}
The linearized Landau operator with Maxwellian molecules
$$\mathscr{L}_Lf=-\mu^{-1/2}Q_L(\mu, \sqrt{\mu}\,f)-\mu^{-1/2} Q_L(\sqrt{\mu}\,f, \mu),$$
acting on the Schwartz space $\mathscr{S}(\rr^d)$ is equal to
\begin{multline*}
\mathscr{L}_L=(d-1)\Big(-\Delta_v+\frac{|v|^2}{4}-\frac{d}{2}\Big)-\Delta_{\SSS^{d-1}}+\Big[\Delta_{\SSS^{d-1}}-(d-1)\Big(-\Delta_v+\frac{|v|^2}{4}-\frac{d}{2}\Big)\Big]\mathbb{P}_1\\ +\Big[-\Delta_{\SSS^{d-1}}-(d-1)\Big(-\Delta_v+\frac{|v|^2}{4}-\frac{d}{2}\Big)\Big]\mathbb{P}_2,
\end{multline*}
where $\Delta_{\SSS^{d-1}}$ stands for the Laplace-Beltrami operator on the unit sphere $\SSS^{d-1}$ and $\mathbb{P}_{k}$ the orthogonal projections onto the Hermite basis.
\end{proposition}

\bigskip

\noindent
We recall that the Laplace-Beltrami operator on the unit sphere $\SSS^{d-1}$ is a sum of squares of vector fields in $\rr^d$ given by the differential operator (see Section~\ref{appendix}),
$$\Delta_{\SSS^{d-1}}=\frac{1}{2}\sum_{\substack{1 \leq j,k \leq d \\ j \neq k}}(v_j \partial_k-v_k \partial_j)^2$$
and that in the 3-dimensional case, the Laplace-Beltrami operator on the unit sphere $\SSS^{2}$ may be considered as a pseudodifferential operator
$$\Delta_{\SSS^{2}}f=(\textrm{Op}^wa)f=\frac{1}{(2\pi)^3}\int_{\rr^{6}}e^{i (v-y) \cdot \xi}a\Big(\frac{v+y}{2},\xi\Big)f(y)dyd\xi,$$
whose Weyl symbol is the anisotropic symbol (see Section~\ref{appendix}),
\begin{equation}\label{symweyl}
a(v,\xi)= \frac{3}{2}-|v \wedge \xi|^2.
\end{equation}
We shall now restrict our study to the three-dimensional setting $d=3$ and recall the definitions of real spherical harmonics.

\noindent
For $\sigma=(\cos \beta \sin \alpha,\sin \beta \sin \alpha,\cos \alpha) \in \SSS^2$ with $\alpha \in [0,\pi]$ and $\beta \in [0,2\pi)$, the real spherical harmonics $Y_l^m(\sigma)$ with $l \in \nn$, $-l \leq m \leq l$, are defined as $Y_0^0(\sigma)=(4\pi)^{-1/2}$ and for any $l \geq 1$,
$$Y_l^m(\sigma)=\begin{cases} \Big(\frac{2l+1}{4\pi}\Big)^{1/2}P_l(\cos \alpha), & \mbox{if } m=0 \\ \Big(\frac{2l+1}{2\pi}\frac{(l-m)!}{(l+m)!}\Big)^{1/2}P_l^m(\cos \alpha)\cos m \beta  & \mbox{if } m=1,...,l \\
\Big(\frac{2l+1}{2\pi}\frac{(l+m)!}{(l-m)!}\Big)^{1/2}P_l^{-m}(\cos \alpha)\sin m \beta  & \mbox{if } m=-l,...,-1,
\end{cases}$$
where $P_l$ stands for the $l$-th Legendre polynomial and $P_l^m$ the associated Legendre functions of the first kind of order $l$ and degree $m$. The family $(Y_l^m)_{l \geq 0, -l \leq m \leq l}$ constitutes an orthonormal basis of the space $L^2(\SSS^2,d\sigma)$ with $d\sigma$ being the surface measure on $\SSS^2$. We set for any $n,l \geq 0$, $-l \leq m \leq l$,
\begin{equation}\label{eig11}
\varphi_{n,l,m}(v)=2^{-3/4}\Big(\frac{2n!}{\Gamma(n+l+\frac{3}{2})}\Big)^{1/2}\Big(\frac{|v|}{\sqrt{2}}\Big)^lL_n^{[l+\frac{1}{2}]}\Big(\frac{|v|^2}{2}\Big)e^{-\frac{|v|^2}{4}}Y_l^m\Big(\frac{v}{|v|}\Big),
\end{equation}
where $L_n^{[l+\frac{1}{2}]}$ are the generalized Laguerre polynomials. The family $(\varphi_{n,l,m})_{n,l \geq 0, |m| \leq l}$ is an orthonormal basis of $L^2(\rr^3)$
composed by eigenvectors of the harmonic oscillator and the Laplace-Beltrami operator on the unit sphere $\SSS^2$,
\begin{equation}\label{k4.1}
\Big(-\Delta_v+\frac{|v|^2}{4}-\frac{3}{2}\Big)\varphi_{n,l,m}=(2n+l)\varphi_{n,l,m}, \quad -\Delta_{\SSS^2}\varphi_{n,l,m}=l(l+1)\varphi_{n,l,m}.
\end{equation}
The space of the collisional invariants (\ref{coli}) may be expressed through this basis as
$$\mathcal{N}= \textrm{Span}\big\{\varphi_{0,0,0},\varphi_{0,1,-1},\varphi_{0,1,0},\varphi_{0,1,1},\varphi_{1,0,0}\big\}.$$
We deduce from Proposition~\ref{th1} and (\ref{k4.1}) that the linearized Landau operator is diagonal in the $L^2(\rr^3)$ orthonormal basis $(\varphi_{n,l,m})_{n,l \geq 0, |m| \leq l}$,
$$\mathcal{L}_{L}\varphi_{n,l,m}=\lambda_{L}(n,l,m)\varphi_{n,l,m}. \quad n,l \geq 0, \ -l \leq m \leq l,$$
where $\lambda_L(0,0,0)=\lambda_L(0,1,0)=\lambda_L(0,1,\pm 1)=\lambda_L(1,0,0)=0, \lambda_L(0,2,m)=12$, and for $2n+l>2$
\begin{equation}\label{k4.22}
\lambda_L(n,l,m)=2(2n+l)+l(l+1).
\end{equation}
We consider now the 3-dimensional Boltzmann collision operator with Maxwellian molecules
$$Q_B(g, f)=\int_{\rr^3}\int_{\SSS^{2}}b\Big(\frac{v-v_*}{|v-v_*|} \cdot \sigma\Big) \big(g'_* f'-g_*f\big)d\sigma dv_*,$$
whose cross section
$$b(\cos \theta)=b\Big(\frac{v-v_*}{|v-v_*|} \cdot \sigma\Big),$$
is supported on the set where $0 \leq \theta \leq \frac{\pi}{2}$ and satisfies to the singularity assumption
\begin{equation}\label{sa1bb}
\sin \theta \ b(\cos \theta)  \substack{\\ \\ \approx \\ \theta \to 0_{+} }  \theta^{-1-2s},
\end{equation}
for some  $0 < s <1$. We refer the reader to Section~\ref{maxboltzmann} for details about the definition of the Boltzmann operator under the singularity assumption (\ref{sa1bb}).
The linearized non-cutoff Boltzmann operator
$$
\mathscr{L}_Bf=-\mu^{-1/2}Q_B(\mu, \sqrt{\mu}\,f)-\mu^{-1/2} Q_B(\sqrt{\mu}\,f, \mu)
$$
is also diagonal in the same orthonormal basis $(\varphi_{n,l,m})_{n,l \geq 0, |m| \leq l}$. In the cutoff case i.e. when
$$
b(\cos \theta)\sin \theta \in L^1([0,\pi/2]),
$$
it was shown in \cite{wang} (see also \cite{bobylev,17,dolera}) that
\begin{equation}\label{k4.23}
\mathscr{L}_{B}\varphi_{n,l,m}=\lambda_{B}(n,l,m)\varphi_{n,l,m}, \quad n,l \geq 0, \ -l \leq m \leq l,
\end{equation}
with
\begin{multline}\label{k4.24}
\lambda_B(n,l,m)=4\pi \int_{0}^{\frac{\pi}{4}}b(\cos 2\theta)\sin(2\theta)\\ \times \big(1+\delta_{n,0}\delta_{l,0}-P_l(\cos \theta)(\cos \theta)^{2n+l}-P_l(\sin \theta)(\sin \theta)^{2n+l}\big)d\theta,
\end{multline}
where $P_l$ are the Legendre polynomials defined by the Rodrigues formula
 \begin{equation}\label{rodrigues}
 P_l(x)=\frac{1}{2^l l!}\frac{d^l}{dx^l}(x^2-1)^l, \quad l \geq 0.
 \end{equation}
By using the properties $P_l(1)=1$, $l \geq 0$ (see e.g. (4.2.7) in \cite{lebedev}) and $P_l(-x)=(-1)^lP_l(x)$, we notice that the smooth function
$$F(\theta)=1+\delta_{n,0}\delta_{l,0}-P_l(\cos \theta)(\cos \theta)^{2n+l}-P_l(\sin \theta)(\sin \theta)^{2n+l},$$
is even and vanishes at zero. It follows from (\ref{sa1bb}) that the function
$$b(\cos 2\theta)\sin(2\theta)F(\theta)=\mathcal{O}(\theta^{1-2s}),$$
when $\theta \rightarrow 0$, is integrable in $0$ and that the integral in (\ref{k4.24}) is also well-defined in the non-cutoff case when the assumption (\ref{sa1bb}) is satisfied. Since the eigenfunctions (\ref{eig11}) are independent on the cross section, we deduce by passing to the limit from the cutoff case to the non-cutoff case
$$\lim_{\substack{\eps \to 0\\ \eps >0}}\un_{[\eps,\frac{\pi}{4}]}(\theta)b(\cos 2\theta)\sin(2\theta) =b(\cos 2\theta)\sin(2\theta),$$
that the diagonalization (\ref{k4.23}) holds true also in the non-cutoff case.
We easily check that the eigenvalues $\lambda_B(n,l,m)$ are all non-negative. Indeed, by using the property (see e.g. (4.4.2) in \cite{lebedev})
\begin{equation}\label{k6.15}
\forall l \geq 0, \forall |x| \leq 1, \ |P_l(x)| \leq 1,
\end{equation}
we have $\lambda_B(0,0,0)=\lambda_B(0,1,\pm 1)=\lambda_B(0,1,0)=0$ and
$$\lambda_B(n,l,m) \geq 4\pi \int_{0}^{\frac{\pi}{4}}b(\cos 2\theta)\sin(2\theta) \underbrace{\big(1-|\cos \theta|^{2n+l}-|\sin \theta|^{2n+l}\big)}_{\geq 0}d\theta \geq 0,$$
when $2n+l \geq 2$.
We recover directly that the two linearized operators $\mathscr{L}_L$ and $\mathscr{L}_B$ are both non-negative
$$(\mathscr{L}_Lf,f)_{L^2} \geq 0, \quad  (\mathscr{L}_Bf,f)_{L^2} \geq 0, \quad f \in \mathscr{S}(\rr^3)$$
and satisfy
$$(\mathscr{L}_Lf,f)_{L^2(\rr^3)}=0 \Leftrightarrow  (\mathscr{L}_Bf,f)_{L^2(\rr^3)}=0 \Leftrightarrow f=\mathbf{P}f,$$
i.e.
$$\lambda_L(n,l,m)=0 \Leftrightarrow \lambda_B(n,l,m)=0.$$
The following result shows that the eigenvalues of the linearized non-cutoff Boltzmann operator $\lambda_B(n,l,m)$ have the same growth as the fractional eigenvalues of the linearized Landau operator $\lambda_L(n,m,l)^s$.

\bigskip

\begin{theorem}\label{th2}
There exists a positive constant $c_0>0$ such that, for any
$n,l \geq 0,\,\,  -l \leq m \leq l$,
$$
\frac{1}{c_0}\big(1+(2n+l)^s+l^s(l+1)^s\big) \leq 1+\lambda_{B}(n,l,m) \leq c_0 \big(1+(2n+l)^s+l^s(l+1)^s\big)
$$
and
$$
\frac{1}{c_0}\lambda_{L}(n,l,m)^s \leq \lambda_{B}(n,l,m) \leq c_0 \lambda_{L}(n,l,m)^s.
$$
\end{theorem}

\bigskip

\noindent
We notice from (\ref{k4.22}) and (\ref{k4.24}) that the eigenvalues $\lambda_L(n,l,m)$ and $\lambda_B(n,l,m)$ depend only on the non-negative parameters $2n+l$, $l(l+1)$, and
from (\ref{k4.1}) that the harmonic oscillator and the Laplace-Beltrami operator commute $[\mathcal{H},\Delta_{\SSS^2}]=0.$
We deduce from Theorem~\ref{th2} that there exists a positive function $\alpha : \nn^2 \rightarrow [c_0^{-1},c_0]$ such that
\begin{equation}\label{eig111}
\forall n,l \geq 0,\forall  -l \leq m \leq l, \quad \lambda_{B}(n,l,m)=\alpha\big(2n+l,l(l+1)\big)\lambda_{L}(n,l,m)^s.
\end{equation}
It therefore follows from (\ref{k4.1}) and (\ref{eig111}) that we can define by the functional calculus the operators $\mathcal{H}$ and $\Delta_{\SSS^2}$,
$$A=a(\mathcal{H},\Delta_{\SSS^2}) : L^2(\rr^3) \rightarrow L^2(\rr^3),$$
a positive bounded isomorphism
$$\exists c>0, \forall f \in L^2(\rr^3), \ c\|f\|_{L^2}^2 \leq \big(a(\mathcal{H},\Delta_{\SSS^2})f,f\big)_{L^2} \leq \frac{1}{c}\|f\|_{L^2}^2$$
 satisfying
$$\mathscr{L}_B=a(\mathcal{H},\Delta_{\SSS^2}) \mathscr{L}_L^s,$$
where the fractional power of the linearized Landau operator is defined through functional calculus. We sum-up these results:

\bigskip

\begin{theorem}
In the case of Maxwellian molecules $\gamma=0$, there exists
$$A=a(\mathcal{H},\Delta_{\SSS^2}) : L^2(\rr^3) \rightarrow L^2(\rr^3),$$
a positive bounded isomorphism
$$\exists c>0, \forall f \in L^2(\rr^3), \ c\|f\|_{L^2}^2 \leq \big(a(\mathcal{H},\Delta_{\SSS^2})f,f\big)_{L^2} \leq \frac{1}{c}\|f\|_{L^2}^2,$$ such that
$$\mathscr{L}_B=a(\mathcal{H},\Delta_{\SSS^2}) \mathscr{L}_L^s.$$
\end{theorem}

\bigskip

\noindent
By using that the Hermite functions are Schwartz functions, we deduce from Proposition~\ref{th1} and (\ref{symweyl}) that the Weyl symbol of linearized Landau operator
$$\mathscr{L}_L=l^w(v,D_v),$$
satisfies
$$l(v,\xi)=2\Big(|\xi|^2+\frac{|v|^2}{4}-\frac{3}{2}\Big)+|v \wedge \xi|^2-\frac{3}{2} \textrm{ mod } S^{-\infty}(\rr^6).$$
Here, we define the symbol classes $\mathbf{S}^m(\rr^{2d})$, for $m \in \rr$, as the set of smooth functions $a(v,\xi)$ from  $\R^d\times \R^d$ into $\C$ satisfying to the estimates
\begin{equation}\label{2.symcl}
\forall (\alpha,\beta) \in \nn^{2d}, \exists C_{\alpha\beta}>0, \forall (v,\xi) \in \rr^{2d},
|\partial_v^{\alpha}\partial_{\xi}^{\beta}a(v,\xi)| \leq C_{\alpha,\beta} \langle (v,\xi) \rangle^{2m-|\alpha|-|\beta|},
\end{equation}
with
$
\valjp{(v,\xi)}=\sqrt{1+\val v^2+\val \xi^2}.
$
The symbol class $\mathbf{S}^{-\infty}(\rr^{2d})$ denotes the class $\cap_{m \in \rr}\mathbf{S}^m(\rr^{2d})$.
We deduce from (\ref{ker}), (\ref{symweyl}) and Theorem~\ref{th2} the following coercive estimates:

\bigskip

\begin{theorem}\label{th4}
In the case of Maxwellian molecules $\gamma=0$, the linearized non-cutoff Boltzmann operator satisfies to the following coercive estimates:
$$(\mathscr{L}_Bf,f)_{L^2}+\|f\|_{L^2}^2 \sim \Big\|\Big(\emph{\textrm{Op}}^w\Big(|\xi|^2+\frac{|v|^2}{4}\Big)\Big)^{\frac{s}{2}}f\Big\|_{L^2}^2 +\big\|\big(\emph{\textrm{Op}}^w(|\xi \wedge v|^2)\big)^{\frac{s}{2}}f\big\|_{L^2}^2, \quad f \in \mathscr{S}(\rr^3)$$
and
\begin{multline*}
\forall f \in \mathscr{S}(\rr^3), \quad   \|\mathcal{H}^{\frac{s}{2}}(1-{\bf P})f\|_{L^2}^2 +\|(-\Delta_{\SSS^2})^{\frac{s}{2}}(1-{\bf P})f\|_{L^2}^2 \\ \lesssim (\mathscr{L}_Bf,f)_{L^2} \lesssim  \|\mathcal{H}^{\frac{s}{2}}(1-{\bf P})f\|_{L^2}^2 +\|(-\Delta_{\SSS^2})^{\frac{s}{2}}(1-{\bf P})f\|_{L^2}^2.
\end{multline*}
\end{theorem}

\bigskip

\noindent
Here the two operators
$$\Big(\textrm{Op}^w\Big(|\xi|^2+\frac{|v|^2}{4}\Big)\Big)^{\frac{s}{2}} \textrm{ and } \big(\textrm{Op}^w(|\xi \wedge v|^2)\big)^{\frac{s}{2}},$$
are defined through functional calculus.
We shall now consider the general three-dimensional case when the molecules are not necessarily Maxwellian, that is, when the parameter $\gamma$ in the kinetic factor (\ref{sa0}) may range over the interval $]-3,+\infty[$. In this case, the linearized non-cutoff Boltzmann operator satisfies to the following weighted coercive estimates:

\bigskip

\begin{theorem}\label{th5}
In the case of general molecules $\gamma \in ]-3,+\infty[$, the linearized non-cutoff Boltzmann operator satisfies to the following coercive estimates:
$$(\mathscr{L}_Bf,f)_{L^2} \sim \Big\|\Big(\emph{\textrm{Op}}^w\Big(|\xi|^2+\frac{|v|^2}{4}\Big)\Big)^{\frac{s}{2}}\langle v \rangle^{\frac{\gamma}{2}}(1-{\bf P})f\Big\|_{L^2}^2 +\big\|\big(\emph{\textrm{Op}}^w(|\xi \wedge v|^2)\big)^{\frac{s}{2}}\langle v \rangle^{\frac{\gamma}{2}}(1-{\bf P})f\big\|_{L^2}^2, \ f \in \mathscr{S}(\rr^3)$$
and
\begin{multline*}
\forall f \in \mathscr{S}(\rr^3), \quad   \|\mathcal{H}^{\frac{s}{2}}\langle v \rangle^{\frac{\gamma}{2}}(1-{\bf P})f\|_{L^2}^2 +\|(-\Delta_{\SSS^2})^{\frac{s}{2}}\langle v \rangle^{\frac{\gamma}{2}}(1-{\bf P})f\|_{L^2}^2 \\ \lesssim (\mathscr{L}_Bf,f)_{L^2} \lesssim  \|\mathcal{H}^{\frac{s}{2}}\langle v \rangle^{\frac{\gamma}{2}}(1-{\bf P})f\|_{L^2}^2 +\|(-\Delta_{\SSS^2})^{\frac{s}{2}}\langle v \rangle^{\frac{\gamma}{2}}(1-{\bf P})f\|_{L^2}^2.
\end{multline*}
\end{theorem}

\bigskip

\noindent
These coercive estimates for general molecules are proven in Section~\ref{nonmax}. They are a direct byproduct of the coercive estimates established in the Maxwellian case (Theorem~\ref{th4}) and the link between Maxwellian and non-Maxwellian cases highlighted in~\cite{amuxy-4-1}.

\section{Proof of the main results}

\subsection{Proof of Proposition~\ref{th1}}

\subsubsection{The linearized operator $\mathscr{L}_{1,L}$}
We consider the first part in the linearized Landau operator
$$\mathscr{L}_{1,L}f=-\mu^{-1/2}Q_L(\mu,\mu^{1/2}f).$$
Let $f \in \mathscr{S}(\rr^d)$ be a Schwartz function. By using that
$$\mu(v_*)\nabla(\mu^{1/2}f)(v)-(\nabla \mu)(v_*)\mu^{1/2}(v)f(v)=\mu(v_*)\mu^{1/2}(v)\Big((\nabla f)(v)-\frac{v}{2}f(v) +v_*f(v)\Big),$$
we have
\begin{multline*}
(\mathscr{L}_{1,L}f)(v)=-\mu^{-1/2}(v)\sum_{1 \leq i,j \leq d} \partial_{v_i}\left(\int_{\RR^d}a_{i,j}(v-v_*)\mu(v_*)\mu^{1/2}(v)\Big(\partial_jf(v)-\frac{v_j}{2}f(v) +v_j^*f(v)\Big)dv_*\right),
\end{multline*}
where we have written $v_*=(v_1^*,...,v_d^*)$.
By using that
\begin{equation}\label{k1.4}
a_{i,i}(v-v_*)=\sum_{\substack{1 \leq k \leq d \\ k \neq i}}(v_k-v_k^*)^2, \quad a_{i,j}(v-v_*)=-(v_i-v_i^*)(v_j-v_j^*) \textrm{ when }  i\neq j, 
\end{equation}
we may write $\mathcal{L}_{1,L}f=A_1f+A_2f$ with
$$(A_1f)(v)=-\mu^{-1/2}(v)\sum_{\substack{1 \leq i,j \leq d \\ i \neq j}} \partial_{v_i}\left(\int_{\RR^d}(v_j-v_j^*)^2\mu(v_*)\mu^{1/2}(v)\Big(\partial_if(v)-\frac{v_i}{2}f(v) +v_i^*f(v)\Big)dv_*\right),$$
$$(A_2f)(v)=\mu^{-1/2}(v)\sum_{\substack{1 \leq i,j \leq d\\ i \neq j}} \partial_{v_i}\left(\int_{\RR^d}(v_i-v_i^*)(v_j-v_j^*)\mu(v_*)\mu^{1/2}(v)\Big(\partial_jf(v)-\frac{v_j}{2}f(v) +v_j^*f(v)\Big)dv_*\right).$$
A direct computation shows that
$$\mathscr{L}_{1,L}f=A_0(v)f+\sum_{j=1}^dB_j(v)\partial_jf+\sum_{\substack{1 \leq i,j \leq d\\ i \neq j}}C_{i,j}(v)\partial_{i,j}^2f+\sum_{j=1}^dD_j(v)\partial_j^2f.$$
The term $A_0$ writes as
$$A_0(v)=\int_{\RR^d}\sum_{\substack{1 \leq i,j \leq d \\ i \neq j}}\tilde{a}_{i,j}(v,v_*)\mu(v_*)dv_*,$$
with
\begin{multline*}
\tilde{a}_{i,j}(v,v_*)=\frac{1}{4} (v_j-v_j^*)^2(-v_i^2 +2v_iv_i^*)+\frac{1}{2}(v_j-v_j^*)^2+\frac{1}{2}(v_j-v_j^*)(-v_j+2v_j^*)\\ -\frac{v_i}{4}(v_i-v_i^*)(v_j-v_j^*)(-v_j+2v_j^*)=\frac{1}{4}\big(2v_jv_j^*-v_i^2v_jv_j^*+v_iv_i^*v_j^2-2(v_j^*)^2+v_i^2(v_j^*)^2-v_iv_jv_i^*v_j^*\big).
\end{multline*}
It follows that
$$A_0(v)=\frac{1}{4}\sum_{\substack{1 \leq i,j \leq d \\ i \neq j}}\int_{\RR^d}\big(-2(v_j^*)^2+v_i^2(v_j^*)^2\big)\mu(v_*)dv_*=(d-1)\frac{|v|^2}{4}-\frac{d(d-1)}{2},$$
since
$$\int_{\rr^d}v_j^*\mu(v_*)dv_*=0, \quad \int_{\rr^d}(v_j^*)^2\mu(v_*)dv_*=1.$$
The term $B_j$ writes as
$$B_j(v)=\int_{\RR^d}b_{j}(v,v_*)\mu(v_*)dv_*,$$
with
\begin{multline*}
b_{j}(v,v_*)=\sum_{\substack{1 \leq i \leq d \\ i \neq j}}(v_i-v_i^*)^2(v_j -v_j^*) + (d-1)(v_j-v_j^*) -\frac{1}{2}\sum_{\substack{1 \leq i \leq d\\ i \neq j}} v_i(v_i-v_i^*)(v_j-v_j^*)\\
+\sum_{\substack{1 \leq i \leq d\\ i \neq j}} (v_j-v_j^*)(v_i-v_i^*)\Big(-\frac{v_i}{2} +v_i^*\Big)=(d-1)(v_j-v_j^*).
\end{multline*}
It follows that $B_j(v)=(d-1)v_j$. When $i \neq j$, the term $C_{i,j}$ writes as
$$C_{i,j}(v)=\int_{\RR^d}(v_i-v_i^*)(v_j-v_j^*)\mu(v_*)dv_*=\int_{\RR^d}(v_iv_j-v_i^*v_j-v_iv_j^*+v_i^*v_j^*)\mu(v_*)dv_*=v_iv_j.$$
The term $D_j$ writes as
$$D_j(v)=-\sum_{\substack{1 \leq i \leq d \\ i \neq j}}\int_{\RR^d}(v_i-v_i^*)^2\mu(v_*)dv_*=-(d-1)-\sum_{\substack{1 \leq i \leq d \\ i \neq j}}v_i^2.$$
We deduce from (\ref{beltra}) that
\begin{multline}\label{k2.6}
\mathscr{L}_{1,L}f=\Big((d-1)\frac{|v|^2}{4}-\frac{d(d-1)}{2}\Big)f+(d-1)\sum_{j=1}^dv_j\partial_jf+\sum_{\substack{1 \leq i,j \leq d\\ i \neq j}}v_iv_j\partial_{i,j}^2f\\ -(d-1)\Delta_vf-\sum_{\substack{1 \leq i,j \leq d \\ i \neq j}}v_i^2\partial_j^2f=(d-1)\Big(-\Delta_v+\frac{|v|^2}{4}-\frac{d}{2}\Big)f-\Delta_{\SSS^{d-1}}f.
\end{multline}

\subsubsection{The linearized operator $\mathscr{L}_{2,L}$}
We consider the second part in the linearized Landau operator
$$\mathscr{L}_{2,L}f=-\mu^{-1/2}Q_L(\mu^{1/2}f,\mu).$$
Let $f \in \mathscr{S}(\rr^d)$ be a Schwartz function. By using that
$$\mu^{1/2}(v_*)f(v_*)(\nabla \mu)(v)-\nabla(\mu^{1/2}f)(v_*)\mu(v)=\mu(v)\mu^{1/2}(v_*)\Big(-f(v_*)v+\frac{v_*}{2}f(v_*)
-(\nabla f)(v_*)\Big),$$
we have $\mathscr{L}_{2,L}f=A_1f+A_2f$ with
\begin{multline*}
(A_1f)(v)=\mu^{-1/2}(v)\int_{\RR^d}\sum_{1 \leq i,j \leq d} \partial_{v_i}\big[a_{i,j}(v-v_*)v_j\mu(v)\big]\mu^{1/2}(v_*)f(v_*) dv_*\\ -\frac{1}{2}\mu^{-1/2}(v)\int_{\RR^d}\sum_{1 \leq i,j \leq d} \partial_{v_i}\big[a_{i,j}(v-v_*)\mu(v)\big]v_j^*\mu^{1/2}(v_*)f(v_*) dv_*,
\end{multline*}
\begin{multline*}
(A_2f)(v)=\mu^{-1/2}(v)\int_{\RR^d}\sum_{1 \leq i,j \leq d} \partial_{v_i}\big[a_{i,j}(v-v_*)\mu(v)\big]\mu^{1/2}(v_*)(\partial_{j} f)(v_*)dv_*\\
=-\mu^{-1/2}(v)\int_{\RR^d}\sum_{1 \leq i,j \leq d} \partial_{v_i,v_j^*}^2\big[a_{i,j}(v-v_*)\mu(v)\mu^{1/2}(v_*)\big]f(v_*)dv_*,
\end{multline*}
by integrating by parts.
We obtain that
$$(\mathscr{L}_{2,L}f)(v)=\int_{\RR^d}K(v,v_*)\mu^{1/2}(v)\mu^{1/2}(v_*)f(v_*) dv_*,$$
where
\begin{multline*}
K(v,v_*)=\sum_{1 \leq i \leq d} a_{i,i}(v-v_*)+\sum_{1 \leq i,j \leq d} \partial_{v_i}\big[a_{i,j}(v-v_*)\big]v_j-\sum_{1 \leq i,j \leq d} a_{i,j}(v-v_*)v_iv_j\\
-\frac{1}{2}\sum_{1 \leq i,j \leq d} \partial_{v_i}\big[a_{i,j}(v-v_*)\big]v_j^*+\frac{1}{2}\sum_{1 \leq i,j \leq d} a_{i,j}(v-v_*)v_iv_j^*-\sum_{1 \leq i,j \leq d} \partial_{v_i,v_j^*}^2\big[a_{i,j}(v-v_*)\big]\\
+\frac{1}{2}\sum_{1 \leq i,j \leq d} \partial_{v_i}\big[a_{i,j}(v-v_*)\big]v_j^*+\sum_{1 \leq i,j \leq d} \partial_{v_j^*}\big[a_{i,j}(v-v_*)\big]v_i-\frac{1}{2}\sum_{1 \leq i,j \leq d} a_{i,j}(v-v_*)v_iv_j^*,
\end{multline*}
that is
\begin{multline*}
K(v,v_*)=\sum_{1 \leq i \leq d} a_{i,i}(v-v_*)+\sum_{1 \leq i,j \leq d} \partial_{v_i}\big[a_{i,j}(v-v_*)\big]v_j-\sum_{1 \leq i,j \leq d} a_{i,j}(v-v_*)v_iv_j\\
-\sum_{1 \leq i,j \leq d} \partial_{v_i,v_j^*}^2\big[a_{i,j}(v-v_*)\big]
+\sum_{1 \leq i,j \leq d} \partial_{v_j^*}\big[a_{i,j}(v-v_*)\big]v_i.
\end{multline*}
By using (\ref{k1.4}), we notice that
$$\sum_{1 \leq i,j \leq d} \partial_{v_i}\big[a_{i,j}(v-v_*)\big]v_j+\sum_{1 \leq i,j \leq d} \partial_{v_j^*}\big[a_{i,j}(v-v_*)\big]v_i=0.$$
We have
\begin{multline*}
K(v,v_*)=\sum_{1 \leq i \leq d} a_{i,i}(v-v_*)-\sum_{1 \leq i,j \leq d} a_{i,j}(v-v_*)v_iv_j
-\sum_{1 \leq i,j \leq d} \partial_{v_i,v_j^*}^2\big[a_{i,j}(v-v_*)\big]\\
=\sum_{1 \leq i \leq d}\sum_{\substack{1 \leq j \leq d \\ i \neq j}}(v_j-v_j^*)^2(1-v_i^2)+\sum_{\substack{1 \leq i,j \leq d\\ i \neq j}}(v_i-v_i^*)(v_j-v_j^*)v_iv_j
-\sum_{\substack{1 \leq i,j \leq d\\ i \neq j}}1.
\end{multline*}
It follows that
\begin{multline*}
K(v,v_*)=(d-1)\sum_{1 \leq j \leq d}\big(v_j^2-2v_jv_j^*+(v_j^*)^2\big)+\sum_{\substack{1 \leq i,j \leq d\\ i \neq j}}(v_j-v_j^*)\big((v_i-v_i^*)v_iv_j -(v_j-v_j^*)v_i^2\big)\\ -d(d-1)=(d-1)\sum_{1 \leq j \leq d}\big(v_j^2-2v_jv_j^*+(v_j^*)^2\big)+\sum_{\substack{1 \leq i,j \leq d\\ i \neq j}}\big(v_iv_i^*v_jv_j^*-v_i^2(v_j^*)^2\big)-d(d-1),
\end{multline*}
because
$$\sum_{\substack{1 \leq i,j \leq d\\ i \neq j}}(v_i^2v_jv_j^*-v_iv_i^*v_j^2)=0.$$
We deduce from (\ref{k2.0}) and (\ref{k2.1}) that
\begin{multline*}
K(v,v_*)\mu^{1/2}(v)\mu^{1/2}(v_*)=-2(d-1)\sum_{1 \leq j \leq d}\Psi_{e_j}(v)\Psi_{e_j}(v_*)+\sum_{\substack{1 \leq i,j \leq d\\ i \neq j}}\Psi_{e_i+e_j}(v)\Psi_{e_i+e_j}(v_*)\\ -2\sum_{\substack{1 \leq i,j \leq d\\ i \neq j}}\Psi_{2e_i}(v)\Psi_{2e_j}(v_*).
\end{multline*}
It follows that
\begin{multline*}
\mathscr{L}_{2,L}f=-2(d-1)\sum_{1 \leq j \leq d}(f,\Psi_{e_j})_{L^2}\Psi_{e_j}+\sum_{\substack{1 \leq i,j \leq d\\ i \neq j}}(f,\Psi_{e_i+e_j})_{L^2}\Psi_{e_i+e_j}-2\sum_{\substack{1 \leq i,j \leq d\\ i \neq j}}(f,\Psi_{2e_i})_{L^2}\Psi_{2e_j}.
\end{multline*}
By using (\ref{beltra}), direct computations provide
$$\Delta_{\SSS^{d-1}}\Psi_{e_j}=-(d-1)\Psi_{e_j}, \quad \Delta_{\SSS^{d-1}}\Psi_{2e_j}=-2(d-1)\Psi_{2e_j}+2\sum_{\substack{1 \leq k \leq d \\ k \neq j}}\Psi_{2e_k},\quad \Delta_{\SSS^{d-1}}\Psi_{e_i+e_j}=-2d\Psi_{e_i+e_j},$$
when $i \neq j$. This implies that
\begin{equation}\label{k2.5}
\mathscr{L}_{2,L}f=\Big[\Delta_{\SSS^{d-1}}-(d-1)\Big(-\Delta_v+\frac{|v|^2}{4}-\frac{d}{2}\Big)\Big]\mathbb{P}_1f\\ +\Big[-\Delta_{\SSS^{d-1}}-(d-1)\Big(-\Delta_v+\frac{|v|^2}{4}-\frac{d}{2}\Big)\Big]\mathbb{P}_2f.
\end{equation}
Then, Proposition~\ref{th1} is a consequence of the identities (\ref{k2.6}) and (\ref{k2.5}).

\subsection{Proof of Theorem~\ref{th2}}
In order to prove Theorem~\ref{th2}, we may assume the cross section satisfies
$$4\pi b(\cos 2\theta)\sin(2\theta)=\frac{1}{\theta^{2s+1}}.$$
When $n+l \geq 1$, we split the term
$$\lambda_B(n,l,m)= \int_{0}^{\frac{\pi}{4}}\frac{1}{\theta^{2s+1}}\big(1-P_l(\cos \theta)(\cos \theta)^{2n+l}-P_l(\sin \theta)(\sin \theta)^{2n+l}\big)d\theta,$$
into three parts
\begin{equation}\label{k7.0}
\lambda_{B}(n,l,m)=\lambda_{1,B}(n,l)+\lambda_{2,B}(n,l)+\lambda_{3,B}(n,l),
\end{equation}
where
$$
\lambda_{1,B}(n,l)=- \int_{0}^{\frac{\pi}{4}}\frac{1}{\theta^{2s+1}}P_l(\sin \theta)(\sin \theta)^{2n+l}d\theta,
$$
\begin{equation}\label{k6.3}
\lambda_{2,B}(n,l)= \int_{0}^{\frac{1}{l+2}}\frac{1}{\theta^{2s+1}}\big(1-P_l(\cos \theta)(\cos \theta)^{2n+l}\big)d\theta > 0
\end{equation}
and
\begin{equation}\label{k6.31}
\lambda_{3,B}(n,l)= \int_{\frac{1}{l+2}}^{\frac{\pi}{4}}\frac{1}{\theta^{2s+1}}\big(1-P_l(\cos \theta)(\cos \theta)^{2n+l}\big)d\theta > 0.
\end{equation}
We recall from (\ref{k6.15}) that the two last terms $\lambda_{2,B}(n,l)$ and $\lambda_{3,B}(n,l)$ are positive when $n+l \geq 1$. The following lemma shows that the first term $\lambda_{1,B}(n,l)$ is exponentially decreasing when $2n+l \rightarrow +\infty$.

\bigskip

\begin{lemma}\label{l1}
We have
$$\forall n,l \geq 0, \ 2n+l \geq 2, \qquad   |\lambda_{1,B}(n,l)| \leq  \frac{1}{2n+l-2s}\Big(\frac{\pi}{4}\Big)^{2n+l-2s},$$
so that $\lambda_{1,B}(n,l)$ is exponentially decreasing when $2n+l \rightarrow +\infty$.
\end{lemma}

\bigskip

\begin{proof}
By using that  $\sin x \leq x$ for $x \geq 0$, we deduce from (\ref{k6.15}) that
$$|\lambda_{1,B}(n,l)| \leq  \int_{0}^{\frac{\pi}{4}}\frac{(\sin \theta)^{2n+l}}{\theta^{2s+1}}d\theta \leq  \int_{0}^{\frac{\pi}{4}}\theta^{2n+l-2s-1}d\theta=\frac{1}{2n+l-2s}\Big(\frac{\pi}{4}\Big)^{2n+l-2s},$$
when $2n+l \geq 2$.
\end{proof}

\bigskip

\noindent
The next lemma provides some estimates for the second term $\lambda_{2,B}(n,l)$:

\bigskip

\begin{lemma}\label{l2}
There exists a positive constant $C>0$ such that
$$\forall n,l \geq 0, \ 2n+l \geq 1, \quad 0 < \lambda_{2,B}(n,l) \leq C \big((2n+l)^s+(l+2)^{2s}\big).$$
Furthermore, we have
$$ \lambda_{2,B}(n,l) \sim (2n+l)^s \int_0^{+\infty}\frac{1}{\theta^{2s+1}}(1-e^{-\frac{\theta^2}{2}})d\theta,$$
when  $\frac{\sqrt{2n+l}}{l+2} \rightarrow +\infty$.
\end{lemma}

\bigskip

\begin{proof}
In order to estimate the term (\ref{k6.3}), we shall be using the Hilb formula \cite{szego} (Theorem~8.21.6),
\begin{equation}\label{hilb}
P_l(\cos \theta)=\Big(\frac{\theta}{\sin \theta}\Big)^{\frac{1}{2}}J_0\Big(\Big(l+\frac{1}{2}\Big)\theta\Big)+\mathcal{O}(\theta^2), \quad l \geq 1,
\end{equation}
when $0<\theta \leq \frac{c}{l}$, where $c>0$ is a fixed constant and $J_0$ the Bessel function of the first kind of order zero
\begin{equation}\label{bessel}
J_0(t)=\frac{1}{\pi}\int_{-\frac{\pi}{2}}^{\frac{\pi}{2}}\cos(t \sin \tau)d\tau.
\end{equation}
We may write
\begin{equation}\label{k6.41}
\lambda_{2,B}(n,l)=\tilde{\lambda}_{2,B}(n,l)+R_{2,B}(n,l),
\end{equation}
with
\begin{equation}\label{k6.42}
\tilde{\lambda}_{2,B}(n,l)=\int_{0}^{\frac{1}{l+2}}\frac{1}{\theta^{2s+1}}\Big[1-\Big(\frac{\theta}{\sin \theta}\Big)^{\frac{1}{2}}J_0\Big(\Big(l+\frac{1}{2}\Big)\theta\Big)(\cos \theta)^{2n+l}\Big]d\theta,
\end{equation}
where the remainder term $R_{2,B}(n,l)$ can be estimated from above as follows
\begin{equation}\label{lel1}
\exists C>0, \forall n,l \geq 0, \ 2n+l \geq 1, \quad |R_{2,B}(n,l)| \leq \frac{C}{(2n+l)^{1-s}}.
\end{equation}
It is sufficient to prove that
$$ \int_{0}^{\frac{1}{l+2}}\frac{(\cos \theta)^{2n+l}}{\theta^{2s-1}}d\theta \lesssim \frac{1}{(2n+l)^{1-s}},$$
when $2n+l \geq 1$. To that end,
we notice that
\begin{multline*}
\int_{0}^{\frac{1}{l+2}}\frac{(\cos \theta)^{2n+l}}{\theta^{2s-1}}d\theta  \leq \int_{0}^{\frac{1}{l+2}}\frac{e^{-\frac{2}{\pi^2}(2n+l)\theta^2}}{\theta^{2s-1}}d\theta\\
 \leq \frac{1}{(2n+l)^{1-s}} \int_{0}^{\frac{\sqrt{2n+l}}{l+2}}\frac{e^{-\frac{2\theta^2}{\pi^2}}}{\theta^{2s-1}}d\theta  \leq \frac{1}{(2n+l)^{1-s}} \int_{0}^{+\infty}\frac{e^{-\frac{2\theta^2}{\pi^2}}}{\theta^{2s-1}}d\theta \lesssim \frac{1}{(2n+l)^{1-s}},
\end{multline*}
because
\begin{equation}\label{k10.0}
\forall \theta \in \Big[0,\frac{1}{2}\Big], \quad (\cos \theta)^{2n+l}=e^{(2n+l)\ln(1-2\sin^2\frac{\theta}{2})} \leq e^{-2(2n+l)\sin^2\frac{\theta}{2}}\leq e^{-\frac{2}{\pi^2}(2n+l)\theta^2}
\end{equation}
since
\begin{equation}\label{k8.01}
\forall \ 0 \leq x<1, \ \ln(1-x) \leq -x, \quad \forall \ 0 \leq x \leq \frac{\pi}{2}, \quad \sin x \geq \frac{2}{\pi}x.
\end{equation}
We shall now study the main contribution in the term (\ref{k6.3}) and prove that
$$\tilde{\lambda}_{2,B}(n,l) \sim (2n+l)^{s}\int_0^{+\infty}\frac{1}{\theta^{2s+1}}(1-e^{-\frac{\theta^2}{2}})d\theta,$$
when  $\frac{\sqrt{2n+l}}{l+2} \rightarrow +\infty$.
We deduce from (\ref{bessel}) and (\ref{k6.42}) that
\begin{multline*}
(2n+l)^{-s}\tilde{\lambda}_{2,B}(n,l)=\\ \frac{1}{\pi}\int_{\theta=0}^{\frac{\sqrt{2n+l}}{l+2}}\int_{\tau=-\frac{\pi}{2}}^{\frac{\pi}{2}}\frac{1}{\theta^{2s+1}}
\Big[1-\Big(\frac{\frac{\theta}{\sqrt{2n+l}}}{\sin \frac{\theta}{\sqrt{2n+l}}}\Big)^{\frac{1}{2}}\cos\Big(\Big(l+\frac{1}{2}\Big)\frac{\theta}{\sqrt{2n+l}}\sin \tau\Big)\Big(\cos \frac{\theta}{\sqrt{2n+l}}\Big)^{2n+l}\Big]d\theta d\tau.
\end{multline*}
Setting
$$F(\theta)=\Big(\frac{\theta}{\sin \theta}\Big)^{\frac{1}{2}},$$
we notice that the even function $F$ is smooth on the interval $[-\frac{\pi}{2},\frac{\pi}{2}]$.
By Taylor expanding the following two terms
$$\cos\Big(\Big(l+\frac{1}{2}\Big)\frac{\theta}{\sqrt{2n+l}}\sin \tau\Big)=1-\Big(l+\frac{1}{2}\Big)^2\frac{\theta^2 \sin^2 \tau}{2n+l}\int_0^1(1-t)\cos\Big(t\Big(l+\frac{1}{2}\Big)\frac{\theta}{\sqrt{2n+l}}\sin \tau\Big)dt,$$
$$\Big(\frac{\frac{\theta}{\sqrt{2n+l}}}{\sin \frac{\theta}{\sqrt{2n+l}}}\Big)^{\frac{1}{2}}=1+\Big(\frac{\theta}{\sqrt{2n+l}}\Big)^2\int_0^1(1-t)F''\Big(t\frac{\theta}{\sqrt{2n+l}}\Big)dt,$$
we may write
\begin{equation}\label{lel2}
(2n+l)^{-s}\tilde{\lambda}_{2,B}(n,l) =A(n,l)+B(n,l)+C(n,l),
\end{equation}
with
$$A(n,l)=\int_{0}^{\frac{\sqrt{2n+l}}{l+2}}\frac{1}{\theta^{1+2s}}\Big[1-\Big(\cos \frac{\theta}{\sqrt{2n+l}}\Big)^{2n+l}\Big]d\theta,$$
\begin{multline*}
B(n,l)=\frac{1}{\pi}\int_{\theta=0}^{\frac{\sqrt{2n+l}}{l+2}}\int_{\tau=-\frac{\pi}{2}}^{\frac{\pi}{2}}\int_{t=0}^1\frac{1}{\theta^{1+2s}}\Big(\frac{\frac{\theta}{\sqrt{2n+l}}}{\sin \frac{\theta}{\sqrt{2n+l}}}\Big)^{\frac{1}{2}}\Big(l+\frac{1}{2}\Big)^2\frac{\theta^2\sin^2 \tau}{2n+l}\\
\times (1-t)\cos\Big(t\Big(l+\frac{1}{2}\Big)\frac{\theta}{\sqrt{2n+l}}\sin \tau\Big)\Big(\cos \frac{\theta}{\sqrt{2n+l}}\Big)^{2n+l}d\theta d\tau dt
\end{multline*}
and
\begin{multline*}
C(n,l)=-\int_{\theta=0}^{\frac{\sqrt{2n+l}}{l+2}}\int_{\tau=-\frac{\pi}{2}}^{\frac{\pi}{2}}\int_{t=0}^1\frac{1-t}{\pi \theta^{1+2s}}\Big(\frac{\theta}{\sqrt{2n+l}}\Big)^2F''\Big(t\frac{\theta}{\sqrt{2n+l}}\Big)\Big(\cos \frac{\theta}{\sqrt{2n+l}}\Big)^{2n+l}d\theta d\tau dt.
\end{multline*}
We deduce from (\ref{k10.0}) and (\ref{k8.01}) that
\begin{equation}\label{lel3}
|B(n,l)| \leq \sqrt{\frac{\pi}{2}}\Big(l+\frac{1}{2}\Big)^2\frac{1}{2n+l}\int_{0}^{\frac{\sqrt{2n+l}}{l+2}}\frac{1}{\theta^{2s-1}}d\theta \leq
\frac{1}{2-2s}\sqrt{\frac{\pi}{2}}\frac{(l+2)^{2s}}{(2n+l)^s}.
\end{equation}
It follows that
\begin{equation}\label{lel3.1}
B(n,l) \rightarrow 0,
\end{equation}
when $\frac{\sqrt{2n+l}}{l+2} \rightarrow +\infty$. Then, we deduce from (\ref{k10.0}) that
\begin{align}\label{lel4}
|C(n,l)| \leq & \  \frac{1}{2n+l}\|F''\|_{L^{\infty}([-\frac{\pi}{2},\frac{\pi}{2}])}\int_{0}^{\frac{\sqrt{2n+l}}{l+2}}\frac{e^{-\frac{2\theta^2}{\pi^2}}}{\theta^{2s-1}}d\theta \\ \leq & \  \frac{1}{2n+l}\|F''\|_{L^{\infty}([-\frac{\pi}{2},\frac{\pi}{2}])}\int_{0}^{+\infty}\frac{e^{-\frac{2\theta^2}{\pi^2}}}{\theta^{2s-1}}d\theta. \notag
\end{align}
It follows that
\begin{equation}\label{lel3.2}
C(n,l) \rightarrow 0,
\end{equation}
when $\frac{\sqrt{2n+l}}{l+2} \rightarrow +\infty$.
Regarding the first term $A(n,l)$, we notice that
$$0 \leq A(n,l) \leq \int_{0}^{+\infty}\frac{1}{\theta^{1+2s}}\Big[1-\Big(\cos \frac{\theta}{\sqrt{2n+l}}\Big)^{2n+l}\Big]d\theta.$$
We notice the pointwise convergence on $]0,+\infty[$,
$$\frac{1}{\theta^{1+2s}} \Big(1-\Big(\cos \frac{\theta}{\sqrt{2n+l}}\Big)^{2n+l}\Big)
\rightarrow \frac{1}{\theta^{2s+1}}(1-e^{-\frac{\theta^2}{2}}),$$
when $2n+l \rightarrow +\infty$. A Taylor expansion shows that for any $\theta>0$,
\begin{align*}
& \  0 \leq  \frac{1}{\theta^{1+2s}} \Big(1-\Big(\cos \frac{\theta}{\sqrt{2n+l}}\Big)^{2n+l}\Big)\\
& \ \ \ \leq   \un_{[1,+\infty[}(\theta)\frac{1}{\theta^{1+2s}} \Big(1-\Big(\cos \frac{\theta}{\sqrt{2n+l}}\Big)^{2n+l}\Big) \\
+  & \ \frac{\un_{]0,1[}(\theta)}{\theta^{2s-1}}\int_0^1(1-t)\Big(\cos\frac{t\theta}{\sqrt{2n+l}}\Big)^{2n+l-2} \Big[\Big(\cos\frac{t\theta}{\sqrt{2n+l}}\Big)^{2}  -(2n+l-1)\Big(\sin\frac{t\theta}{\sqrt{2n+l}}\Big)^{2}\Big]dt \\
 & \ \ \  \leq  \frac{2}{\theta^{1+2s}}\un_{[1,+\infty[}(\theta)+\frac{1}{\theta^{2s-1}}\un_{]0,1[}(\theta)\int_0^1\Big(\cos\frac{t\theta}{\sqrt{2n+l}}\Big)^{2n+l}dt  \\
 & \ \ \ \leq  \frac{2}{\theta^{1+2s}}\un_{[1,+\infty[}(\theta)+\frac{1}{\theta^{2s-1}}\un_{]0,1[}(\theta) \in L^1(]0,+\infty[),
\end{align*}
when $2n+l \geq1$. We deduce from the Lebesgue dominated convergence theorem that
$$\lim_{2n+l \to +\infty}\int_{0}^{+\infty}\frac{1}{\theta^{1+2s}}\Big[1-\Big(\cos \frac{\theta}{\sqrt{2n+l}}\Big)^{2n+l}\Big]d\theta=\int_0^{+\infty}\frac{1}{\theta^{2s+1}}(1-e^{-\frac{\theta^2}{2}})d\theta.$$
It follows that there exists a positive constant $C>0$ such that
\begin{equation}\label{lel6}
\forall n,l \geq 0, \quad 0 \leq A(n,l) \leq C.
\end{equation}
Furthermore, we notice the pointwise convergence
$$\un_{]0,\frac{\sqrt{2n+l}}{l+2}]}(\theta)\frac{1}{\theta^{1+2s}} \Big(1-\Big(\cos \frac{\theta}{\sqrt{2n+l}}\Big)^{2n+l}\Big)
\rightarrow \un_{]0,+\infty[}(\theta)\frac{1}{\theta^{2s+1}}(1-e^{-\frac{\theta^2}{2}}),$$
when $\frac{\sqrt{2n+l}}{l+2} \rightarrow +\infty$. Another use of the Lebesgue dominated convergence theorem shows that
\begin{equation}\label{lel7}
A(n,l) \rightarrow \int_0^{+\infty}\frac{1}{\theta^{2s+1}}(1-e^{-\frac{\theta^2}{2}})d\theta,
\end{equation}
when  $\frac{\sqrt{2n+l}}{l+2} \rightarrow +\infty$. We deduce from (\ref{lel1}), (\ref{lel2}), (\ref{lel3.1}), (\ref{lel3.2}) and (\ref{lel7}) that
$$\lambda_{2,B}(n,l) \sim (2n+l)^{s}\int_0^{+\infty}\frac{1}{\theta^{2s+1}}(1-e^{-\frac{\theta^2}{2}})d\theta,$$
when  $\frac{\sqrt{2n+l}}{l+2} \rightarrow +\infty$. Furthermore, we easily notice from  (\ref{k6.41}), (\ref{lel1}), (\ref{lel2}), (\ref{lel3}), (\ref{lel4}) and (\ref{lel6}) that
$$\exists C>0, \forall n,l \geq 0, \ 2n+l \geq 1, \quad 0 < \lambda_{2,B}(n,l) \leq C \Big((2n+l)^s+\frac{1}{(2n+l)^{1-s}}+(l+2)^{2s}\Big).$$
This ends the proof of Lemma~\ref{l2}
\end{proof}

\noindent
The next lemma provides some upper and lower estimates for the last term (\ref{k6.31}).

\bigskip

\begin{lemma}\label{l4}
There exists a positive constant $c>0$ such that
$$\forall n,l \geq 0, \quad c l^{2s} \leq \lambda_{3,B}(n,l)  \leq \frac{1}{c} (1+l)^{2s}.$$
\end{lemma}

\bigskip

\begin{proof}
We deduce from (\ref{k6.15}) that for all $l \geq 2$,
\begin{multline*}
l^{2s}\int_{1}^{\frac{\pi}{2}}\frac{1}{\theta^{1+2s}}\Big(1-\Big|P_l\Big(\cos \frac{\theta}{l}\Big)\Big|\Big)d\theta \\ \leq \lambda_{3,B}(n,l)= l^{2s}\int_{\frac{l}{l+2}}^{\frac{\pi}{4}l}\frac{1}{\theta^{1+2s}}\Big(1-P_l\Big(\cos \frac{\theta}{l}\Big)\Big(\cos \frac{\theta}{l}\Big)^{2n+l}\Big)d\theta
\leq l^{2s}\int_{\frac{1}{3}}^{+\infty}\frac{2d\theta}{\theta^{1+2s}}.
\end{multline*}
We deduce from the Hilb formula (\ref{hilb}) the pointwise convergence
$$\lim_{l \rightarrow +\infty}P_l\Big(\cos \frac{\theta}{l}\Big)=J_0(\theta).$$
We notice from the definition that the Bessel function $J_0$ is a smooth bounded in modulus by 1 but not identically equal to 1 in modulus on the interval $[1,\frac{\pi}{2}]$.
We also notice from (\ref{rodrigues}) and (\ref{k6.15}) that the smooth function
$$\theta \mapsto P_l\Big(\cos \frac{\theta}{l}\Big),$$
is bounded in modulus by 1 but not identically equal to 1 in modulus on the interval $[1,\frac{\pi}{2}]$ when $l \geq 1$ so that
$$\int_{1}^{\frac{\pi}{2}}\frac{1}{\theta^{1+2s}}\Big(1-\Big|P_l\Big(\cos \frac{\theta}{l}\Big)\Big|\Big)d\theta>0.$$
We deduce from (\ref{k6.15}) and the Lebesgue dominated convergence Theorem that
$$\lim_{l \rightarrow +\infty}\int_{1}^{\frac{\pi}{2}}\frac{1}{\theta^{1+2s}}\Big(1-\Big|P_l\Big(\cos \frac{\theta}{l}\Big)\Big|\Big)d\theta=\int_{1}^{\frac{\pi}{2}}\frac{1}{\theta^{1+2s}}(1-|J_0(\theta)|)d\theta>0,$$
because
$$\forall l \geq 2, \quad \Big|1-\Big|P_l\Big(\cos \frac{\theta}{l}\Big)\Big|\Big| \leq 2.$$
When $l=0$, we deduce from (\ref{k6.15}) that for all $n \geq 0$,
$$0 \leq \lambda_{3,B}(n,0) \leq \int_{\frac{1}{2}}^{\frac{\pi}{4}}\frac{2}{\theta^{2s+1}}d\theta.$$
When $l=1$, we deduce from (\ref{k6.15}) that for all $n \geq 0$,
$$0< \int_{\frac{1}{3}}^{\frac{\pi}{4}}\frac{1}{\theta^{2s+1}}(1-\cos \theta)d\theta \leq \lambda_{3,B}(n,1) \leq \int_{\frac{1}{3}}^{\frac{\pi}{4}}\frac{2}{\theta^{2s+1}}d\theta.$$
This ends the proof of Lemma~\ref{l4}.
\end{proof}

\bigskip

\noindent
Theorem~\ref{th2} is a direct consequence of (\ref{k4.22}), (\ref{k7.0}) and Lemmas~\ref{l1}, \ref{l2}, \ref{l4}.

\subsection{Proof of Theorem~\ref{th5}}\label{nonmax}
We first consider the case with Maxwellian molecules $\gamma=0$. We deduce from (\ref{tripleest}) and Theorem~\ref{th4} the equivalence of the norms
$$\triple f\triple_{0}^2 \sim \Big\|\Big(\textrm{Op}^w\Big(|\xi|^2+\frac{|v|^2}{4}\Big)\Big)^{\frac{s}{2}}f\Big\|_{L^2}^2 +\big\|\big(\textrm{Op}^w(|\xi \wedge v|^2)\big)^{\frac{s}{2}}f\big\|_{L^2}^2 \sim \|\mathcal{H}^{\frac{s}{2}}f\|_{L^2}^2 +\|(-\Delta_{\SSS^2})^{\frac{s}{2}}f\|_{L^2}^2.$$
On the other hand, the following equivalence between the norm $\triple \cdot \triple_0$ in the Maxwellian case and the norm $\triple \cdot \triple_{\gamma}$ for general molecules $\gamma \in ]-3,+\infty[$ was proven in \cite{amuxy-4-1} (Proposition~2.4):
$$\triple f\triple_{\gamma} \sim \triple \langle v \rangle^{\frac{\gamma}{2}}f \triple_0.$$
For general molecules, we therefore obtain that  the linearized non-cutoff Boltzmann operator satisfies to the following coercive estimates:
\begin{align*}
(\mathscr{L}_Bf,f)_{L^2} \sim & \  \triple (1-{\bf P}) f\triple_{\gamma}^2 \sim \triple \langle v \rangle^{\frac{\gamma}{2}}  (1-{\bf P}) f \triple_0\\
 \sim & \  \Big\|\Big(\textrm{Op}^w\Big(|\xi|^2+\frac{|v|^2}{4}\Big)\Big)^{\frac{s}{2}}\langle v \rangle^{\frac{\gamma}{2}} (1-{\bf P})f\Big\|_{L^2}^2 +\big\|\big(\textrm{Op}^w(|\xi \wedge v|^2)\big)^{\frac{s}{2}}\langle v \rangle^{\frac{\gamma}{2}} (1-{\bf P})f\big\|_{L^2}^2  \\ \sim & \  \|\mathcal{H}^{\frac{s}{2}}\langle v \rangle^{\frac{\gamma}{2}} (1-{\bf P})f\|_{L^2}^2 +\|(-\Delta_{\SSS^2})^{\frac{s}{2}}\langle v \rangle^{\frac{\gamma}{2}}  (1-{\bf P})f\|_{L^2}^2.
\end{align*}
This ends the proof of Theorem~\ref{th5}.

\section{Miscellanea}

\subsection{The non-cutoff Boltzmann operator}\label{maxboltzmann}
We consider the non-cutff Boltzmann operator with Maxwellian molecules
$$Q_B(g, f)=\int_{\rr^d}\int_{\SSS^{d-1}}b\Big(\frac{v-v_{*}}{|v-v_*|} \cdot \sigma\Big) \big(g'_* f'-g_*f\big)d\sigma dv_*,$$
with $d \geq 2$, where $f'_*=f(v'_*)$, $f'=f(v')$, $f_*=f(v_*)$, $f=f(v)$. The post collisional velocities are defined in terms of the pre collisional velocities as
$$v'=\frac{v+v_*}{2}+\frac{|v-v_*|}{2}\sigma,\quad   v_*'=\frac{v+v_*}{2}-\frac{|v-v_*|}{2}\sigma,$$
where $\sigma\in\SSS^{d-1}$. We recall here how the Boltzmann operator is defined when the cross section satisfy the singularity assumption (\ref{sa1b}).
To that end, we shall use the distribution of order 2 defined in the following lemma:

\bigskip

\begin{lemma}\label{new003}
Let $\nu$ be an even $L^1_{loc}({\rr^*})$ function satisfying $\theta^2\nu(\theta)\in L^1(\rr)$. Then the mapping
$$
C^2_{c}(\rr)\ni \phi\mapsto\lim_{\varepsilon\rightarrow 0_{+}}\int_{\vert\theta\vert\ge \varepsilon}
\nu(\theta)\bigl(\phi(\theta)-\phi(0)\bigr) d\theta=\int_{0}^1\int_{\rr}\theta^2\nu(\theta)\phi''(t\theta) d\theta (1-t)dt,
$$
is defining a distribution $\finp{(\nu)}$
of order 2. Furthermore, the linear form
$\finp{(\nu)}$ can be extended to $C^{1,1}$ functions
($C^1$ functions whose second derivative is $L^\io$).
For $\phi\in C^{1,1}$ such that
 $\phi(0)=0$, the function $\nu \breve\phi$ belongs to $L^1(\R)$ and
$$\poscal{\finp{(\nu)}}{\phi}=\int\nu(\theta)\breve{\phi}(\theta)d\theta,$$
if $\breve{\phi}$ stands for the even part of the function $\phi$.
\end{lemma}

\bigskip

\begin{proof}
Since
$$\int_{\vert\theta\vert\ge \varepsilon}
\nu(\theta)\bigl(\phi(\theta)-\phi(0)\bigr) d\theta=\int_{0}^1\int_{\vert\theta\vert\ge \varepsilon}\theta^2\nu(\theta)\phi''(t\theta) d\theta (1-t) dt,$$
the Lebesgue dominated convergence theorem gives the first result.
The extension to $C^{1,1}$ functions follows from the formula
$$
\frac12(\phi(\theta)-\phi(0))+
\frac12(\phi(-\theta)-\phi(0))=\frac12\int_{0}^\theta\bigl(\phi'(\tau)-\phi'(-\tau)
\bigr) d\tau,
$$
since the absolute value of the latter is bounded from above by $\norm{\phi''}_{L^\io}\frac{\theta^2}{2}$. This implies that
$$\nu(\theta)\times\text{\tt even part}(\phi(\theta)-\phi(0))\in L^1,$$
and proves the last statement.
\end{proof}

\bigskip

\noindent
By using polar coordinates $v-v_{*}=\rho\nu,\ \rho>0,\ \nu\in \mathbb S^{d-1}$, we may write
$$Q_B(g, f)=\int_{\R_{\rho}^+\times\SSS^{d-1}_{\sigma}\times\SSS^{d-1}_{\nu}}b(\nu\cdot \sigma)\Bigl[g\Big(v-\frac{\rho(\sigma+ \nu)}{2}\Big)f\Big(v+\frac{\rho(\sigma- \nu)}{2}\Big)-g(v-{\rho\nu})f(v)\Bigr]\rho^{d-1}d\sigma d\rho d\nu.$$
Setting $\sigma=\omega\sin \theta\oplus\nu\cos \theta,\ \omega\in \SSS^{d-2},\ \omega\perp \nu,\  0<\theta<\pi,$ the term $Q_B(g, f)$ is equal to
\begin{align*}
& \ \int_{\R_{\rho}^+\times\SSS^{d-2}_{\omega}\times (0,\pi)\times\SSS^{d-1}_{\nu}
}b(\cos \theta) \rho^{d-1} (\sin \theta)^{d-2}
\\&\
\times \Bigl[g\Big(v-\frac{\rho(\omega\sin \theta\oplus\nu\cos \theta+ \nu)}{2}\Big)
f\Big(v+\frac{\rho(\omega\sin \theta\oplus\nu\cos \theta- \nu)}{2}\Big)-
g(v-{\rho\nu})f(v)
\Bigr]d\rho  d\theta d\omega d\nu
=\\  & \
\int_{\R_{\rho}^+\times\SSS^{d-2}_{\omega}\times (0,\pi)\times\SSS^{d-1}_{\nu}
}b(\cos \theta) \rho^{d-1} (\sin \theta)^{d-2}
\\&\
\times \Bigl[g\Big(v-{\rho\cos\frac{\theta}{2}\Big(\omega\sin\frac{\theta}{2}\oplus\nu\cos\frac{\theta}{2}\Big)}\Big)
f\Big(v+{\rho\sin\frac{\theta}{2}\Big(\omega\cos\frac{\theta}{2}\ominus\nu \sin\frac{\theta}{2}\Big)}\Big)-
g(v-{\rho\nu})f(v)
\Bigr]d\theta d\rho d\omega d\nu.
\end{align*}
By using that the cross section $b(\cos \theta)$ is supported on the set where $0 \leq \theta \leq \frac{\pi}{2}$, we have
\begin{multline*}
Q_B(g, f)=\int_{\R_{\rho}^+\times\SSS^{d-2}_{\omega}\times (0,\pi/4)\times\SSS^{d-1}_{\nu}
}2 \rho^{d-1} b(\cos 2\theta)(\sin 2\theta)^{d-2} \\ \times
\Bigl[g\bigl(v-{\rho\cos\theta(\omega\sin\theta\oplus\nu\cos\theta)}\bigr)
f\bigl(v+{\rho\sin\theta(\omega\cos\theta\ominus\nu\sin\theta)}\bigr)
-
g(v-{\rho\nu})f(v)
\Bigr]d\theta d\rho d\omega d\nu.
\end{multline*}
Setting
\begin{multline*}
\Psi_{f,g}(\theta,v)=\int_{\SSS^{d-2}_{\omega}\times \R^+_{\rho}\times\mathbb S^{d-1}_{\nu}}
\hspace{-1.8cm}g\bigl(v-{\rho\cos\theta(\omega\sin\theta\oplus\nu\cos\theta)}\bigr)
f\bigl(v+{\rho\sin\theta(\omega\cos\theta\ominus\nu\sin\theta)}\bigr) \rho^{d-1} d\rho d\omega
d\nu,
\end{multline*}
we notice that the smooth function $\theta\mapsto\Psi_{f,g}(\theta,v)$ is even
\begin{multline*}
\Psi_{f,g}(-\theta,v)=\int_{\SSS^{d-2}_{\omega}\times \R^+_{\rho}\times\mathbb S^{d-1}_{\nu}}\hspace{-1.2cm}
g\bigl(v-{\rho\cos\theta(-\omega\sin\theta\oplus\nu\cos\theta)}\bigr)f\bigl(v-{\rho\sin\theta(\omega\cos\theta\oplus\nu\sin\theta)}\bigr) \rho^{d-1} d\rho
d\omega d\nu
\\ =\int_{\SSS^{d-2}_{\omega}\times \R^+_{\rho}\times\mathbb S^{d-1}_{\nu}} \hspace{-1.2cm}
g\bigl(v-{\rho\cos\theta(\omega\sin\theta\oplus\nu\cos\theta)}\bigr)f\bigl(v+{\rho\sin\theta(\omega\cos\theta\ominus\nu\sin\theta)}\bigr) \rho^{d-1} d\rho
d\omega d\nu=\Psi_{f,g}(\theta,v)
\end{multline*}
and that
$$\Psi_{f,g}(0,v)=\int_{\SSS^{d-2}_{\omega}\times \R^+_{\rho}\times\mathbb S^{d-1}_{\nu}}g(v-\rho\nu)f(v)\rho^{d-1} d\rho d\omega d\nu.$$
For $f,g\in \mathscr{S}(\R^d)$, we easily check that the function $\R^d\ni v\mapsto\p_{\theta}^m\Psi_{f,g}(\theta,v)$ belongs to the Schwartz space $\mathscr S(\R^d)$ uniformly with respect to the parameter $\theta \in (0,\frac{\pi}{4})$ since
\begin{multline*}
\val{v'_{*}}^2+\val{v'}^2=
\Big|v-\rho\cos\frac{\theta}{2}\Big(\omega\sin\frac{\theta}{2}\oplus\nu\cos\frac{\theta}{2}\Big)\Big|^2
+\Big|v+\rho\sin\frac{\theta}{2}\Big(\omega\cos\frac{\theta}{2}\ominus\nu\sin\frac{\theta}{2}\Big)\Big|^2
\\
=2\val v^2+\rho^2-2\rho v\cdot \nu=\val v^2+\val{v-\rho\nu}^2=
\val{v_{*}}^2+\val{v}^2
\ge \frac{1}{3}(\val{v}^2+\rho^2).
\end{multline*}
When the cross section satisfies the assumption (\ref{sa1b}), the Boltzmann operator is then defined as a finite part by Lemma \ref{new003},
$$Q_B(g,f)(v)=\int_{0}^{\frac{\pi}{4}}2  b(\cos 2\theta)(\sin 2\theta)^{d-2} \left(\Psi_{f,g}(\theta,v)-\Psi_{f,g}(0,v) \right)d\theta.$$
Furthermore, we check that $Q_B(g,f)\in \mathscr S(\R^d)$ when $f,g\in \mathscr S(\R^d)$.

\subsection{The Laplace-Beltrami operator on the unit sphere $\SSS^{d-1}$}\label{appendix}
The Laplace-Beltrami operator on the unit sphere $\SSS^{d-1}$ is given by the differential operator
\begin{equation}\label{beltra}
\Delta_{\SSS^{d-1}}=\frac{1}{2}\sum_{\substack{1 \leq j,k \leq d \\ j \neq k}}(v_j \partial_k-v_k \partial_j)^2.
\end{equation}
Indeed, we have
\begin{align*}
& \ \frac{1}{2}\sum_{\substack{1 \leq j,k \leq d \\ j \neq k}}(v_j \partial_k-v_k \partial_j)^2=
\frac{1}{2}\sum_{\substack{1 \leq j,k \leq d \\ j \neq k}}(v_j^2 \partial_k^2+v_k^2 \partial_j^2-2v_jv_k\partial_{j,k}^2-v_j\partial_j-v_k\partial_k)\\
=& \ r^2\Delta_{\rr^d}-\sum_{j=1}^dv_j^2\partial_j^2-\sum_{\substack{1 \leq j,k \leq d \\ j \neq k}}v_jv_k\partial_{j,k}^2-(d-1)\sum_{j=1}^dv_j\partial_j\\
= & \ r^2\Delta_{\rr^d}-\Big(\sum_{j=1}^dv_j\partial_j\Big)^2-(d-2)\sum_{j=1}^dv_j\partial_j
=r^2\Delta_{\rr^d}-(r\partial_r)^2-(d-2)r\partial_r=\Delta_{\SSS^{d-1}},
\end{align*}
because the Laplacian on $\rr^d$ writes in spherical coordinates as
$$\Delta_{\rr^d}=\frac{\partial^2}{\partial r^2}+\frac{d-1}{r}\frac{\partial}{\partial r}+\frac{1}{r^2}\Delta_{\SSS^{d-1}}.$$
In the 3-dimensional case, the Laplace-Beltrami operator on the unit sphere $\SSS^{2}$ is a differential operator on $\rr^3$,
$$\Delta_{\SSS^{2}}f=(\textrm{Op}^wa)f=\frac{1}{(2\pi)^3}\int_{\rr^{6}}e^{i (v-y) \cdot \xi}a\Big(\frac{v+y}{2},\xi\Big)f(y)dyd\xi,$$
whose Weyl symbol is the anisotropic symbol
$$a(v,\xi)= \frac{3}{2}-|v \wedge \xi|^2.$$
Indeed, we have for any $j \neq k$
\begin{multline*}
\textrm{Op}^w\big((v_j\xi_k-v_k\xi_j)^2\big)=\textrm{Op}^w\big(v_j^2\xi_k^2+v_k^2\xi_j^2-2v_j\xi_jv_k\xi_k\big)=-v_j^2\partial_k^2-v_k^2\partial_j^2\\
+\frac{1}{2}(v_j\partial_j+\partial_j v_j)(v_k\partial_k+\partial_k v_k)=-v_j^2\partial_k^2-v_k^2\partial_j^2+v_j\partial_j+v_k\partial_k
+2v_jv_k\partial^2_{j,k}+\frac{1}{2} =\frac{1}{2}-(v_j\partial_k-v_k\partial_j)^2,
\end{multline*}
implying that
$$\Delta_{\SSS^2}=\textrm{Op}^w\Big(\frac{3}{2}-|v \wedge \xi|^2\Big).$$

\bigskip

\vs\noindent
{\bf Acknowledgements.}
The research of the second author was supported by the Grant-in-Aid for Scientific Research No.22540187, Japan Society of the Promotion of Science. The research of the third author was supported by the CNRS chair of excellence at Cergy-Pontoise University. The research of the last   author was supported partially by ``The Fundamental Research Funds for Central Universities''.

\end{document}